%% file: main.tex
\documentclass[conference]{IEEEtran}
\IEEEoverridecommandlockouts

\usepackage{cite}
\usepackage{enumitem}
\usepackage{algorithm}
\usepackage{algpseudocode}
\usepackage{amsmath}
\usepackage{amssymb}
\usepackage{caption}
\usepackage{balance}
\usepackage{multirow} 
\usepackage{graphicx} 
\usepackage{tikz}
\usepackage{pgfplots}
\pgfplotsset{compat=1.18}
\usetikzlibrary{patterns,patterns.meta} 
\usepackage{subfigure}
\definecolor{nord9}{HTML}{81A1C1} 
\definecolor{nord14}{HTML}{A3BE8C}
\usepackage{comment}
\usepackage{url}
\usepackage[most]{tcolorbox}
\usepackage{enumitem}
\usepackage{booktabs}
\usepackage{mathtools}
\usepackage{array}
\newcommand{\ours}{\ensuremath{{\sf DACER}}}

\def\BibTeX{{\rm B\kern-.05em{\sc i\kern-.025em b}\kern-.08em
    T\kern-.1667em\lower.7ex\hbox{E}\kern-.125emX}}
\begin{document}

\title{\Large \bf A Runtime Decentralized Attestation and Coordinated Repair Framework for Securing Automotive ECUs}

\author{
\IEEEauthorblockN{Josh Dafoe}
\IEEEauthorblockA{
\textit{Department of Computer Science} \\
\textit{Michigan Technological University} \\
Houghton, Michigan, USA \\
jwdafoe@mtu.edu
}
\and
\IEEEauthorblockN{Niusen Chen}
\IEEEauthorblockA{
\textit{Department of Computer Science and Engineering} \\
\textit{University of Nevada, Reno} \\
Reno, Nevada, USA \\
niusenc@unr.edu
}
\and
\IEEEauthorblockN{Bo Chen}
\IEEEauthorblockA{
\textit{Department of Computer Science} \\
\textit{Michigan Technological University} \\
Houghton, Michigan, USA \\
bchen@mtu.edu
}
}

\maketitle

\pagestyle{plain}
\thispagestyle{plain}
\input{Source/0-abstract}


\input{Source/1-introduction}
\input{Source/2-background}
\input{Source/3-model}

\input{Source/4-design}
\input{Source/7-discussion}

\input{Source/6-evaluation}
\input{Source/8-related}

\section{Conclusion}
In this work, we have introduced $\ours$, the first framework that integrates malware \textit{detection} and \textit{repair} for ECUs while meeting strict runtime vehicle constraints. Specifically, $\ours$ co-designs efficient attestation and repair mechanisms that unify the ``local'' nature of rollback with the ``global'' nature of ECU reboot. Experimental evaluation of embedded systems in the real-world demonstrates the low overhead of $\ours$.

\section*{Acknowledgment}
This material is based upon work supported by the National Science Foundation Graduate Research Fellowship Program under Grant No.~2437847, and by the National Science Foundation under Grant Nos.~2225424, 2043022 and 2623031. Any opinions, findings, and conclusions or recommendations expressed in this material are those of the author(s) and do not necessarily reflect the views of the National Science Foundation.


\bibliographystyle{plain}
\bibliography{refs}

\appendix
\input{Source/appendix2}
\end{document}

%% file: Source/0-abstract.tex
\begin{abstract}
The evolution of automotive technology increasingly integrates components, transforming vehicles into interconnected systems of systems. Modern vehicles are controlled by a distributed system of computing devices, known as electronic control units (ECUs). However, this interconnectedness means that any error poses significant risks to the vehicle operator. In particular, malware can be injected into ECUs, threatening vehicle safety. To address this, we need mechanisms to detect compromised ECUs then repair them to a benign state. Existing approaches mainly focus on detection and do not address the challenge of integrating detection with runtime ECU repair. This integration is nontrivial because runtime repair involves both local rollback and reboot with timing determined from global vehicle context to avoid unsafe behavior.

In this work, we have designed DACER, a runtime decentralized attestation and coordinated repair framework for automotive ECUs. DACER is the first approach that co-designs attestation and repair to unify the ``local'' nature of firmware rollback with the ``global'' nature of ECU reboot. In DACER, each ECU performs efficient local self-attestation and self-repair functions, enabling low-overhead coordination for distributed operations. In addition, DACER takes advantage of the hierarchical vehicle computing architecture. Our resulting DACER design checks the entire state of the vehicle, resists single points of failure, conforms to real-time constraints, and enables firmware restoration during runtime. The key functions are enabled by the ARM TrustZone equipped within each ECU and the secure flash memory controller embedded in the storage device. We implemented DACER on real-world hardware and experimentally demonstrated its low overhead.

\end{abstract}

%% file: Source/1-introduction.tex
\section{Introduction}

The modern vehicle is increasingly intelligent, with its functions (e.g., accelerating, braking, shifting gears) controlled by specialized computers known as electronic control units (ECUs). These ECUs are resource-constrained, low-power embedded systems that communicate over in-vehicle networks. The vehicle also has a main controller that acts as the ``brain'' of the entire vehicle. Increasingly, the main controller and ECUs are more complex in their software code base. For example, the software of a vehicle can contain up to 100 million lines of code~\cite{antinyan2020revealing}. 
This growing complexity of the software stack has created opportunities for various malware attacks on vehicles. One of the most significant being the injection of malware into vehicle ECUs: previous work has demonstrated that ECU programmability can often be enabled without proper authentication~\cite{kulandaivel2024candid, lauser2023formal}, and the attacker can utilize such vulnerabilities to inject malware into vehicle ECUs in the real world~\cite{valasek2015remote,2023fuzzy}. 

Defending against malware injection attacks on ECUs requires mechanisms that can both detect compromised ECU firmware and repair it. Importantly, the \textit{detection} and \textit{repair} components should be highly integrated because: 1) the repair component needs to be triggered by the detection component, which requires communicating malware detection results to the repair component despite disrupted communications due to compromised firmware, and 2) the repair component should return the compromised ECU to a state that can be verified by the detection component. 

\noindent\textbf{Why is it challenging to integrate the two components in vehicles?} The runtime ECU firmware repair can be decomposed into rollback and reboot: rollback is a restoration of the previous benign firmware in the flash storage, while reboot removes the malware from volatile memory and loads the  restored benign firmware. Hence, a reboot should always be performed \textit{after} the rollback. 
Fundamentally, the rollback is a \textit{local} operation performed by the flash storage of each individual ECU~\cite{guan2017supporting}. Therefore, it does not immediately disrupt ECU functionality and can be performed promptly after detecting malware. Unlike rollback, rebooting an ECU temporarily disrupts its functionality, which can be dangerous depending on the current situation of the vehicle, e.g., rebooting a brake controller during driving should be avoided. Therefore, a reboot decision cannot be made locally and requires a broader vehicle context (i.e., \textit{a context-aware reboot}). 

A typical detection mechanism is firmware attestation~\footnote{Another approach is through intrusion detection~\cite{hoppe2009applying}, which does not directly verify firmware integrity and suffers from a few limitations, for example, malicious firmware that does not produce anomalous network behavior may remain undetected; in addition, a small false positive rate will incorrectly trigger frequent system restorations~\cite{kaster2024automotive}.}
which can be broadly categorized as self-attestation and remote-attestation. We argue that regardless of which detection strategy is used, integrating both detection and repair components to meet critical vehicle constraints is always challenging:

Strategy I: \textit{self-attestation}. In this strategy, each ECU has a trust anchor (e.g. a bootloader loaded from ROM or a trusted execution environment) internally, which will perform firmware verification. 
For a failing verification, it can invoke a local rollback during runtime~\cite{dafoe2025hardware}, but it is hard to coordinate externally for the context-aware reboot due to the potential denial of service (DoS) attacks performed by malicious firmware that controls the network interface. 
Additionally, the existing self-attestation design inherently lacks robustness, as it may be disabled by a compromised or a failed ECU.

Strategy II: \textit{remote-attestation}. In this strategy, a remote entity is designated to verify the integrity of the individual ECU or the integrity of the entire vehicle, and this remote entity can be within the vehicle itself (e.g., a special ECU or multiple ECUs) or outside of the vehicle (e.g., another vehicle or an edge server). This strategy performs an explicit vehicle-wide integrity check, making context-aware reboot possible. However, it
cannot securely invoke the rollback that occurs locally in the compromised ECU due to the potential DoS attack from the malicious firmware. 

\noindent\textbf{How to integrate the two components while meeting vehicle-specific constraints?} A fundamental challenge not addressed by all previous works is to unify local rollback with the external context-aware reboot. To address this challenge, we have proposed $\ours{}$, a \textbf{D}ecentralized \textbf{A}ttestation and \textbf{C}oordinated \textbf{E}CU \textbf{R}epair framework for autonomous vehicles. Our key insight is that locally enforced rollback can be unified with external context-aware reboot by \textit{externalizing self-attestation results}. 
Within each ECU, a failed attestation result triggers timely local rollback. Thus, the externalized results provide other in-vehicle computers with a reliable view of the ECU state.
This information can be used to coordinate the reboot timing among in-vehicle computers with contextual awareness.
As a result, $\ours$ can resolve the shortcomings of the existing approaches~\cite{anand2024achieving,gui2018hardware,sanwald2019secure,kim2017secure,ep3923168b1,oguma2008new,kohnhauser2019ensuring,dafoe2025hardware} by simultaneously
i) complying with strict real-time constraints in the vehicle, ii) verifying the state of the software of the entire vehicle, iii) remaining robust to single point failure for malware detection, and iv) supporting the \textit{safe} automatic runtime restoration of compromised ECUs. 

\ours{} consists mainly of two components: 1) a ``horizontal'' decentralized attestation that identifies malicious ECUs, and 2) a ``vertical'' coordinated repair that enables restoration of compromised ECUs at runtime:

In the \underline{decentralized attestation component}, each of the individual ECUs performs self-attestation locally, and, meanwhile, they coordinate using a trusted execution environment (which also performs self-attestation) to verify whether each local self-attestation state is valid or not. Because the coordinated verification checks only the self-attestation status (which is small in size compared to the ECU firmware image), each decentralized attestation can be performed in constant time and requires only a constant persistent state per node. This low overhead is crucial for the vehicle context which has strict real-time constraints~\cite{cardenas2009challenges}. Additionally, all ECUs can be verifiers, achieving robustness against node failure. 

The \underline{coordinated repair component} uses vehicle-wide integrity information produced through decentralized attestation to guide repair. While local rollback can restore benign firmware on external storage, rebooting an ECU during runtime may be unsafe~\cite{autosar_fota, andreasson2022device}. 
Therefore, $\ours$ securely propagates the attestation results ``up'' to the main controller, which uses its broader view of the vehicle state to determine a safe time to initiate the reboot. Since intermediate nodes may be compromised, $\ours$ applies decentralized attestation across all layers, ensuring that compromised intermediate nodes can be detected and repaired.

\noindent\textbf{Contributions}. Our contributions are summarized below:

\begin{enumerate}[label=(\roman*)]
    \item We have designed the first framework that 1) integrates detection and repair components to address malware injection attacks on ECUs, 2) complies with the critical constraints of vehicles during runtime, and 3) resists against single points of failure.
    \item We have introduced a new decentralized attestation through which the ECUs mutually check the reliable firmware state information produced by self-attestation.
  Then, these results inform our coordinated repair mechanism, in which we utilize the hierarchical nature of the vehicle computing architecture, introducing ``vertical'' coordination between the ECUs and the main controller to make informed decisions on the reboot timing.
    \item We have implemented a prototype of $\ours{}$ in real-world embedded systems and evaluated its performance.
\end{enumerate}

%% file: Source/2-background.tex
\section{Background}
\label{sec:background}

\noindent \textbf{ARM TrustZone (TZ)}. In automotive platforms, in-vehicle computers commonly use ARM-based processors (e.g., Cortex-A and Cortex-M families)~\cite{nxp_s32g3_2025,nxp_s32k3_2025}, where TrustZone support is widely available~\cite{arm_trustzone_cortex_a_2025,arm_trustzone_cortex_m_2025}. TrustZone is a trusted execution environment (TEE) that isolates protected application data and instructions from the untrusted system at the hardware level. In TrustZone, the protected environment is known as the secure world, in which trusted applications are isolated at the hardware level, while the unprotected environment is known as the normal world, which runs the normal host OS. The TrustZone hardware enforces that the normal world cannot access the isolated memory of the secure world, but the secure world can access both normal and isolated memory.
A TrustZone-based application is commonly structured as a trusted component (TApp) running in the secure world and an accompanying untrusted component (UApp) running in the normal world that mediates interaction with the rest of the system.
Calls into the TApp are typically initiated by UApp via a narrow predefined interface (i.e., an SMC instruction). TrustZone also typically supports a mechanism to store secrets that must persist over time (e.g., cryptographic keys) with confidentiality and integrity.

\noindent\textbf{In-Vehicle computing architecture}. Electronic control units (ECUs) are low-power embedded systems that run real-time operating systems. Physically connected ECUs communicate directly over a shared CAN bus, which is a broadcast based communication bus.
The groups of ECUs are known as zones and each zone is managed by a zone controller. Zone controllers typically have more powerful computing hardware and manage the information flow with the rest of the vehicle. Zone controllers typically communicate with each other and a main controller using Ethernet. The main controller synthesizes information and makes high-level control decisions. The control signals are then passed to the correct ECUs through the respective zone controller.

\noindent\textbf{Self-attestation and self-repair.} \textit{Self-attestation} enables a trusted entity within a node to verify software integrity. While traditional secure boot relies on a secure bootloader or TPM check a cryptographic signature over the entire firmware before booting, the strict real-time in-vehicle constraints make this impractical: ECUs are typically required to boot within 50–100 ms~\cite{kim2017secure, ep3923168b1}. Thus, we focus on \textit{probabilistic} self-attestation schemes which reduce the auditing time significantly, although detecting corruptions with higher probability requires multiple attestations over time. 
The trusted component (e.g. an HSM or TEE) achieves this by creating unique challenges on the firmware image, each one corresponding to a fixed-length subset of the firmware~\cite{dafoe2025hardware, nasser2019accelerated,kaster2021sliced}. In this paper, we model the behavior of a probabilistic self-attestation scheme over time by using an epoch time $T_{SA}$ in which exactly one self-attestation is performed, and an associated probability function $\Pr[D \mid \Gamma]$ denoting the probability that corruption is detected in a given attestation, where $\Gamma$ refers to system states (i.e., the proportion of the firmware image that has been modified, the number of attestations already performed, etc.).

\textit{Self-repair} enables the results from an attestation to transition the ECU \textit{storage} into a ``repaired'' state. Previous work proposes a protocol between TrustZone and trusted local storage hardware, allowing an invalid self-attestation result to trigger rollback by a storage controller that maintains the previous benign firmware~\cite{dafoe2024enabling}. Notably, this mechanism ensures rollback to the most recent benign firmware upon either (i) an invalid self-attestation result or (ii) denial of service attacks performed by a Dolev-Yao adversary on the channel between the trusted component and the storage device. In practice, rollback can be achieved using the out-of-place update feature inherent in flash memory storage~\cite{guan2017supporting,dafoe2024enabling} or using a dual-bank update design~\cite{nxp2021an13497}.

%% file: Source/3-model.tex
\section{Models and Assumptions}
\label{sec:models}

\begin{figure}[tb]
  \centering
  \includegraphics[scale=.17]{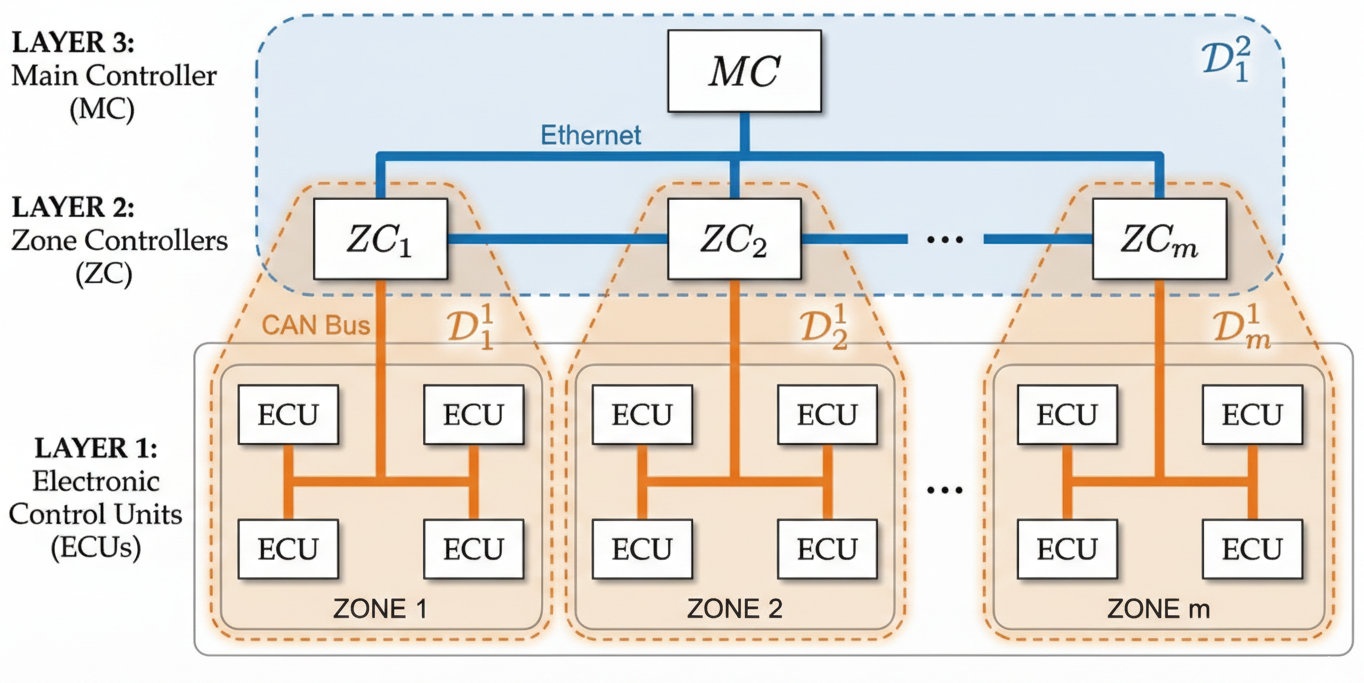}
  \caption{System model overview. The dotted sections represent horizontal domains. The ECUs within a zone are interconnected by CAN bus, and the zone controllers and the main controller are interconnected by Ethernet cable. Note that each zone controller is part of two domains, and acts as a communication gateway between the ECUs and \texttt{MC}.}
  \vspace{-10pt}
  \label{fig:system-model}
\end{figure}

\begin{figure}[tb]
  \centering
  \includegraphics[scale=.11]{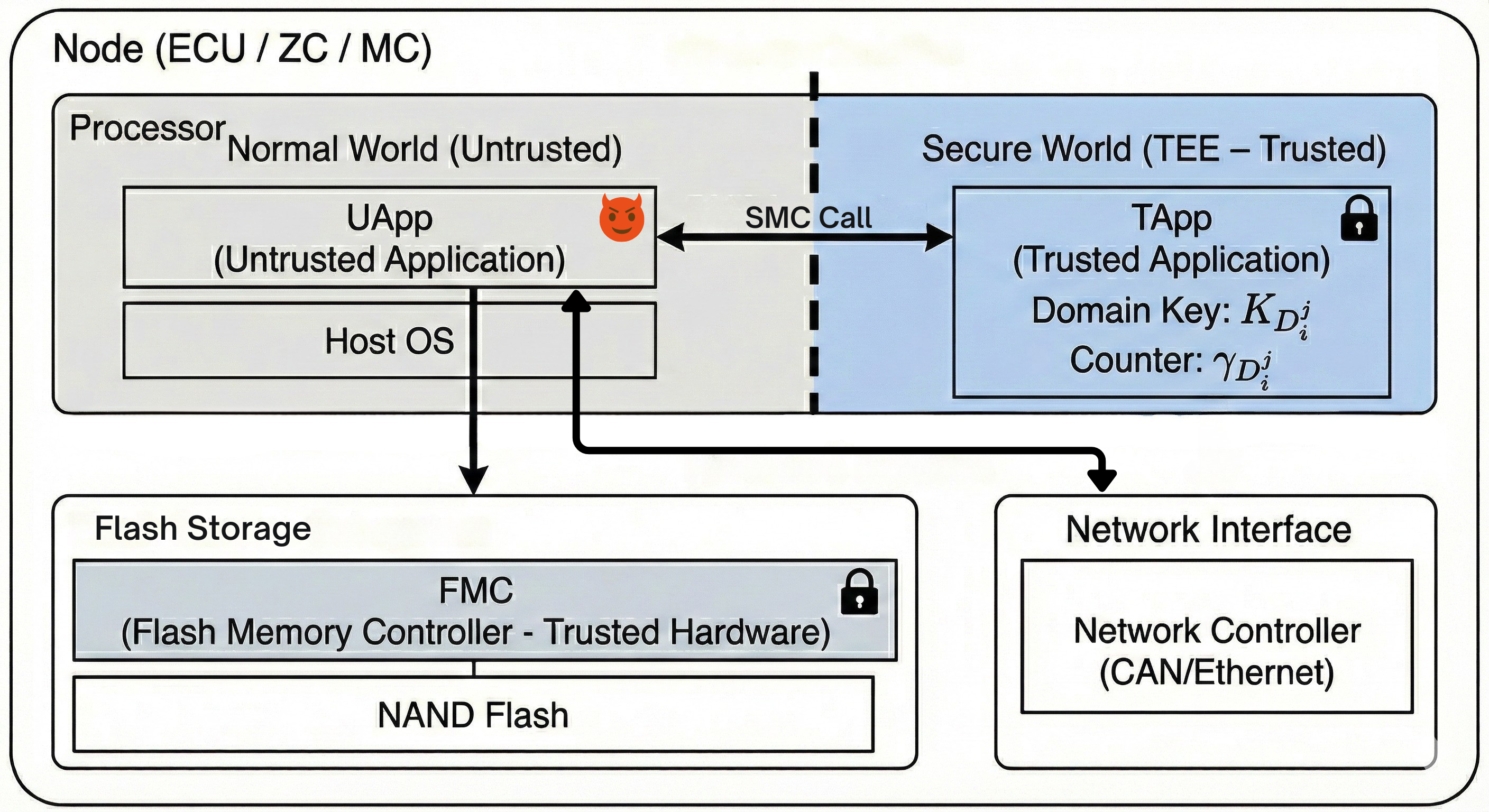}
  \caption{The internal architecture of an in-vehicle computing node (ECU, zone controller, or main controller).}
  \vspace{-10pt}
  \label{fig:system-model-node}
\end{figure}
\noindent\textbf{System model}. 
We consider a modern vehicle following a layered architecture (\S~\ref{sec:background}). Specifically, we consider $3$ layers, each consisting of one or more in-vehicle computers (Figure~\ref{fig:system-model}). Layer $3$ contains only the main controller (\texttt{MC}). Layer $2$, in turn, contains $m$ zone controllers, each acting as the parent node for a zone of ECUs. As such, each zone controller can communicate directly with all ECUs in its zones, other zone controllers, and the main controller. Then, the ECUs make up Layer $1$ and are partitioned into $m$ disjoint \textit{zones}, each corresponding to a zone controller (each zone is managed by one zone controller). Within a zone, the ECUs can communicate directly, while the ECUs in different zones cannot communicate directly; this communication must be facilitated by the zone controllers. Importantly, we define a horizontal \textit{domain} as a group of peers that can communicate directly (e.g. the ECUs in a single zone, along with their parent zone controller). We use $D^\ell_d$ to denote the $d$th horizontal domain associated with layer $\ell$. In our three-layer architecture, Layer $1$ has $m$ horizontal domains, $D^1_1,\ldots,D^1_m$, where each domain consists of the ECUs in one zone together with their zone controller. Layer $2$ has a single horizontal domain, $D^2_1$, consisting of all $m$ zone controllers and \texttt{MC}. Our design can be easily extended to support additional layers.

The internal architecture of an in-vehicle computer is shown in Figure~\ref{fig:system-model-node}, which is a typical computing system equipped with a processor, RAM, and flash memory. Flash memory is managed by a flash memory controller (\texttt{FMC}) that runs secure firmware isolated from the host OS at the hardware level. Importantly, this secure firmware performs some mechanism sufficient to enable a self-repair primitive \cite{dafoe2025hardware, nxp2021an13497}.
Additionally, each processor is enabled with TrustZone~\footnote{Some in-vehicle computer (such as the main controller) may use a different TEE. Our design can be adapted if that TEE meets assumptions 2 and 3 in the ``assumptions'' of~\S~\ref{sec:models}.}. In the secure world, a trusted application (\texttt{TApp}) is running, and in the normal world, an untrusted application (\texttt{UApp}) is running. Unlike \texttt{TApp}, which is isolated in the secure world, \texttt{UApp} which runs on the host OS can directly access the \texttt{FMC} read/write interface and the network controllers (i.e. CAN controller or Ethernet controller). Therefore, communications between \texttt{TApp} and any other component go through \texttt{UApp}~\footnote{Another option is to add storage/network driver functionality into \texttt{TApp}. However, this would not be practical because it 1) requires significant implementation effort, and 2) increases the overall memory footprint, which is expensive for the resource-constrained ECUs.}. 
For each domain, the manufacturer generates a key $K_{D_d^\ell}$ and counter $\gamma_{D_d^\ell}$ and shares them between \texttt{TApp} in all nodes in the domain.
Additionally, the local self-attestation and self-repair functions may use a shared key between \texttt{TApp} and \texttt{FMC} (unique to each ECU). Unless otherwise noted, all counters are incremented after each use.

\noindent\textbf{Adversarial model}. An attacker may compromise the firmware of any ECU. In particular, if the adversary can gain access to the in-vehicle network (e.g., through OTA update paths, diagnostic tooling, compromised OBD-II devices, or compromised mobile phones~\cite{bielawski2020FirmwareUpdates}), 
they may invoke the ECU’s reprogramming mode and install malicious firmware. Note that we only consider an attacker that aims to establish persistence, and so we consider the malware which is committed to flash memory. This injected malware may disrupt the ECU function or send malicious control messages over the CAN bus. Furthermore, since \texttt{UApp} mediates all communication between \texttt{TApp} and \texttt{FMC}, and between \texttt{TApp} and the CAN bus, a malicious \texttt{UApp} may perform man-in-the-middle (MITM) attacks on these communication channels. We model this as a Dolev-Yao-style adversary, positioned between \texttt{TApp} and both the \texttt{FMC} and the network interfaces. This adversary can observe, forge, modify, delay, or drop messages on these channels. Additionally, there may be malware on any in-vehicle computer.

\noindent\textbf{Assumptions}. Our design relies on a few assumptions: 

1) The TrustZone is secure (i.e. \texttt{TApp} is trusted). This is a common assumption in secure designs that rely on secure hardware~\cite{guan2017supporting}. Hardening TrustZone security itself has been actively investigated in the literature~\cite{wan2020rustee}, and is beyond our scope. In addition, the FMC is secure (i.e. \texttt{FMC} is trusted). This assumption can be realized, since i) the \texttt{FMC} firmware is stored in ROM or can be measured via secure boot, and ii) is isolated by the flash storage hardware, which offers a limited read/write interface to the host OS. 
2) For each domain, \texttt{TApp} on the associated parent controller is equipped with a secure mechanism to send a reboot signal to each computer in its domain. A ``secure'' mechanism to send a reboot signal is defined as a signal that can be sent only by the \texttt{TApp}, guaranteeing that the target ECU will reboot. This mechanism can be implemented as a GPIO pin on the parent, accessible only by \texttt{TApp}~\cite{stmicroelectronics2021an5600}, which is connected to the reset pin of the target ECU. 3) There is a direct physical communication interface between the vehicle operator and \texttt{TApp} in \texttt{MC}. This interface can be implemented through GPIO pins~\cite{stmicroelectronics2021an5600} that are only accessible to \texttt{TApp} in \texttt{MC}.
4) The clocks of all in-vehicle computers maintain adequate synchronization using standard automotive network protocols~\cite{autosar_timesync_can_2023}.

%% file: Source/4-design.tex
\section{\ours{} Design}
\label{sec:design}

\subsection{Design Overview}

$\ours$ is organized into two primary functional components: decentralized attestation and coordinated repair. Decentralized attestation is a protocol for detecting malware in ECUs. Each ECU runs a local self-attestation, which allows other
nodes to efficiently check the software integrity of the ECUs in their domain. Coordinated repair then leverages the self-attestation results to first perform local rollback, then uses the decentralized-attestation results to inform context-aware repair decisions on reboot timing of the target nodes, enabling safe restoration at runtime.

To enable external awareness of an ECU's firmware integrity, $\ours$ performs a decentralized attestation, in which the nodes in a domain collaborate to verify each other's firmware integrity.
An immediate solution is to directly employ traditional remote-attestation, where each node directly verifies a cryptographic proof. However, this is expensive as it requires computation over the entire firmware image to generate a proof, and maintaining per-node state information (e.g. hashes) to verify. More importantly, traditional remote-attestation only enables external verification of firmware integrity. Because the verification result is produced outside the compromised ECU, it cannot reliably trigger the local firmware rollback needed for runtime repair. $\ours$ unifies the local nature of rollback and distributed nature of reboot by adopting a probabilistic self-attestation primitive on each ECU, and developing a decentralized-attestation protocol for externalizing the self-attestation results. To achieve this, we leverage TrustZone as the trusted component responsible for both local self-attestation verification and decentralized coordination among nodes within the same domain. Our decentralized-attestation protocol enforces that nodes are periodically challenged to report their state to a set of verifiers.

After self-attestation results trigger rollback via self-repair, the ECU needs to be rebooted in a timely manner.
If at least one verifier in the decentralized-attestation protocol is benign, then the coordinated-repair can be initialized (see~\S~\ref{sec:discussion}). While the latest benign firmware has already been restored to local storage, the code executing from RAM remains malicious until it reboots. The coordinated repair helps ensure that this reboot is safe and secure. Specifically, the reboot timing should not be decided locally, as there is limited contextual information and a runtime reboot is not always safe~\cite{andreasson2022device}. For example, the vehicle may attempt to stop while the brake controller is rebooting, causing a collision. 
We propose that the decision about when to reboot a malicious ECU for repair should be propagated ``up'' to \texttt{MC}, which has the necessary contextual information and computational capabilities to determine a safe reboot window. Importantly, while the decentralized attestation is ``horizontal'' within a domain, the coordinated repair is ``vertical'', since the reboot signal is initiated above the target node. However, this vertical approach introduces a dependency on the parent zone controller, since the communication path used to repair an ECU must remain reliable. We address this problem by applying $\ours$ at multiple layers. Specifically, the zone controllers participate in a horizontal domain with \texttt{MC}, allowing malicious zone controllers to be detected by their peers and directly repaired by \texttt{MC}. Recovery of \texttt{MC} itself relies on Assumption~3. Then, once the zone controllers are repaired, they can correctly forward the detection alert to further repair the ECUs.

\subsection{Decentralized Attestation}

To properly update a target ECU, an external entity should initiate a reprogramming mode. Importantly, the reprogramming mode typically allows arbitrary manipulation of the ECU firmware. The reprogramming mode is entered over CAN through a protocol like Unified Diagnostic Services (UDS). This update mechanism may be compromised by an attacker, who may upload malware to an ECU~\cite{herrewegen2018beneath, kulandaivel2024candid, lauser2023formal}.

Therefore, it is necessary to check whether the firmware running on a given ECU is provided by a valid software distributor (such as the OEM). While self-attestation can perform this check locally, coordinated repair requires that result to be visible to other in-vehicle computers. To externalize this local integrity state, \ours{} leverages decentralized attestation among nodes in the same horizontal domain. Within each domain, a random node is periodically challenged and any node in the domain can verify this response.

The remaining challenge is how a node can externally verify the local self-attestation state of another node without performing the expensive firmware measurement used by traditional remote-attestation. To address this challenge, we decouple the decentralized attestation protocol from firmware measurement. This firmware measurement is performed only by probabilistic self-attestation. Then, a decentralized attestation is reduced to verifiers asking the question: have all local self-attestations on the challenged ECU returned valid results since the last reboot?
To answer this question, the challenged ECU TrustZone releases a secret verifiable ``challenge value'' only when its self-attestation state is valid. Importantly, this value depends only on per-challenge state derived from the shared domain key ($K_{D_d^\ell}$), not on any per-ECU firmware measurement. This information is independent of the ECU being challenged and enables any node in a domain to verify any other node with constant storage regardless of the domain's size. The remainder of this section describes our protocol for secure verification of this information.

Similarly to self-attestation, decentralized attestation challenges are issued periodically. Due to the substantial difference in computing power between in-vehicle computers at different layers, we may desire different epoch lengths for different layers. We define $T_{\ell}$ as the time length of a decentralized attestation epoch in domains at layer $\ell$. In our protocol, periodic challenges are generated locally by \texttt{TApp} in each node. This eliminates any communication and coordination overhead associated with distributing challenges and agreeing on challenge parameters for each epoch. During each epoch, one node is challenged (this may be extended to more nodes, as discussed in \S~\ref{sec:discussion}) by one or more verifiers, and the challenge consists of (i) the challenged node ID, (ii) a set of verifier IDs, and (iii) the challenge value. Based on (i) and (ii), \texttt{TApp} in the challenged node and verifier peers can identify themselves and respond appropriately. The purpose of (iii) is to act as a secret per-challenge value that only the trusted component in each node can produce. 

To generate challenges, respond to them, and verify the response, we rely on a few secrets shared between the \texttt{TApp}s. In each domain, we rely on the key $K_{D_d^\ell}$ and counter $\gamma_{D_d^\ell}$ shared between \texttt{TApp} in each node in the domain. In addition, we define $l$, which is a global security parameter, and $v_d^\ell$ which is the number of verifiers per epoch in a given domain. Importantly, the above shared secrets constitute the only state which needs to be maintained in each in-vehicle computer to achieve the decentralized attestation (enabling mutual verification with constant per-node state, avoiding the pairwise state of alternative designs).

In addition, we define a pseudo random function with variable output length, $g: \{0,1\}^{\kappa} \times \{0,1\}^* \times \mathbb{N} \rightarrow \{0,1\}^*$ such that for every key $K \in \{0,1\}^{\kappa}$, input $x \in \{0,1\}^*$, and length $L \in \mathbb{N}$, we have $g(K,x,L) \in \{0,1\}^L$. The details of challenge generation procedure are presented in Algorithm~\ref{alg:challenge-generation}.

\begin{figure}[tb]
\begin{algorithm}[H]  
\caption*{\textbf{Algorithm 1} Decentralized Attestation Challenge Generation}
\begin{algorithmic}[1]  
\Require Key $K$, counter $\gamma$, number of nodes $m$, number of verifiers $v$, response length $l$, current node index $i$.
\Ensure Role of current node $R$, challenge value $C$ (if applicable), challenged node index $p$, verifier indices $V$.
\Procedure{GenerateChallenge}{$K, \gamma, m, v, l, i$}  
\State $p \gets g(K, \gamma, \lceil \log_2 m \rceil) \bmod m$  
\State $V \gets \emptyset$ 
\For{$t \gets 0$  $v-1$}  
\State $k \gets g(K, \gamma, \lceil \log_2 m \rceil) \bmod m$ 
\State $V.\mathsf{append}(k)$
\EndFor  
\State $C \gets g(K, \gamma, l)$
\If{$i = p$}  
  \State \Return $(\mathsf{Challenged}, C, p, V)$  
\ElsIf{$i \in V$}  
  \State \Return $(\mathsf{Verifier}, C, p, V)$  
\Else  
  \State \Return $(\mathsf{None}, C, p, V)$
\EndIf  
\EndProcedure
\end{algorithmic}  
\end{algorithm}
\captionof{algorithm}{Decentralized Attestation Challenge Generation. Each invocation of $g$ increments $\gamma$ (i.e., $\gamma \leftarrow \gamma + 1$), so that all nodes have the same input, and derive identical $(p,V,C)$ without coordination. When constructing $V$, duplicates are removed and resampled, and the challenged ID $p$ is excluded. Additionally, only nodes with roles \textsf{Challenged} or \textsf{Verifier} use $C$ (others may compute and discard it to preserve $\gamma$ synchronization).}
\label{alg:challenge-generation}
\vspace{-15pt}
\end{figure}

At the beginning of each epoch, every node will run Algorithm~\ref{alg:challenge-generation}, with the correct input associated with its domain. This is used for \texttt{TApp} in each node to 1) generate the challenge value $C$, and 2) determine whether the given node is challenged, is a verifier, or neither. The decentralized-attestation protocol works as follows:

First, in the challenged node, \texttt{TApp} reveals the challenge value $C$ to \texttt{UApp} if and only if the node's local self-attestation state is valid. \texttt{UApp} is then expected to broadcast $C$ to the verifier nodes. However, since \texttt{UApp} is untrusted, it broadcasts some value $C'$, which may differ from $C$ either because \texttt{UApp} never received $C$, or because \texttt{UApp} tampered with it. Note that, since self-attestation is probabilistic, there is a small probability (i.e. $1 - Pr[D \mid \Gamma]$) that the challenged node is compromised, yet \texttt{UApp} received $C$.

Then, each verifier node's \texttt{UApp} either receives some value $C'$ or receives no response before the timeout. If a value is received, the verifier's \texttt{UApp} forwards it to its local \texttt{TApp}. Since the verifier's \texttt{UApp} is also untrusted, it may forward some value $C''$ (which may or may not equal $C'$) instead.

Finally, the verifier \texttt{TApp} compares $C''$ with the challenge value $C$ that it independently generated. If $C''$ equals $C$, then the challenged node is considered valid for that epoch. Otherwise, either the challenged node failed to produce the correct response, or the verifier's \texttt{UApp} is malicious and manipulated $C'$. To distinguish these cases, the verifier \texttt{TApp} checks its own self-attestation state. If the verifier has failed any self-attestations, it cannot conclude that the challenged node definitely contains malware, and therefore suppresses the report. If the verifier's self-attestation state is valid, then the challenged node has malware unless the verifier is itself compromised and undetected, which occurs with probability at most $1 - \Pr[D \mid \Gamma]$. The verifier \texttt{TApp} therefore generates an invalid attestation report.

\subsection{Coordinated Repair}

Upon external detection of malware via decentralized attestation, it is highly likely that the challenged node has failed its internal self-attestation, as analyzed in \S~\ref{sec:discussion}. Consequently, a rollback has been performed (or is imminent \S~\ref{sec:background}). However, as previously mentioned, this rollback only addresses persistent storage, and the malware remains active in the node's volatile memory.
Therefore, a safe and timely reboot is required. To achieve this, the verifier nodes will propagate an invalid attestation report ``up'' to \texttt{MC}, which can orchestrate a safe reboot of the target ECU. 

The first challenge towards propagating this attestation report is that, in malicious nodes, the adversary may manipulate communications between \texttt{TApp} and the CAN bus. Since \texttt{UApp} can simply drop the invalid attestation report, we can only assure that benign peers will actually propagate the message. Therefore, ruling out false-negatives requires a benign node to be part of the verifier set. The remaining challenge is to prevent the adversary from forging false positives.

To achieve this, \texttt{TApp} computes a MAC over the invalid attestation report, by generating 
$$
M = \text{``invalid''}||g(K_{D_d^\ell},\gamma_{D_d^\ell} || p || \text{``invalid''},l)
$$. Importantly, the counter $\gamma_{D_d^\ell}$ referenced in this section is \textit{not} incremented during the generation or verification of these messages. This is because this computation takes place only within a subset of the verifier nodes for this epoch, and incrementing $\gamma_{D_d^\ell}$ would create an inconsistency across nodes. This is justified since the message is bound to the \emph{current} $\gamma_{D_d^\ell}$ value, which will be advanced during the next synchronized challenge generation step (i.e., at the next epoch transition). Replays \emph{within} the same epoch can be suppressed by de-duplicating reports per target node at the receiver \texttt{TApp} (e.g., \texttt{MC}).

$M$ will then be sent, through \texttt{UApp}, to the parent node. A benign parent node will definitely forward this to \texttt{TApp}, but a compromised parent may drop $M$. If the parent is not \texttt{MC}, then $M$, in this form, cannot be verified by \texttt{MC}, since \texttt{MC} is not in the same domain as the target ECU. Therefore, \texttt{MC} does not have access to $K_{D_d^\ell}$ or $\gamma_{D_d^\ell}$. However, \texttt{MC} is in the same domain as the zone controller, and so the zone controller can generate a new MAC, and forward this to \texttt{MC}.

Then, \texttt{TApp} in \texttt{MC} will verify $M$, and make a decision about when to safely reboot the compromised node, based on the rich contextual information available only to \texttt{MC}. When the decision is made to reboot the compromised node, this signal is propagated down, which is feasible by our assumption 2. Importantly, while a zone controller may contain malware at some time (breaking the dependency chain for coordinated repair), assumption 3 ensures that \texttt{MC} can always be repaired, which can eventually repair the zone controllers.

%% file: Source/7-discussion.tex
\section{Analysis and Discussion}
\label{sec:discussion}

\subsection{Security Analysis}

To analyze the security of decentralized attestation, we consider false positives, false negatives, and eventual recovery in the worst case. A more detailed security analysis including a derivation of the mean time to repair is given in~\S~\ref{sec:security:repair}.

\noindent \textit{For false positives}, we observe that this can only occur when $C' = C$ and $C'' \neq C$ for some verifier ECU. In addition, this ECU will only generate a false positive if it has \textit{not} failed any self-attestations. Given $\zeta$ malicious verifiers, where the malware in verifier $k$ is independently detected via
spot-checking with probability $\Pr[D_k \mid \Gamma]$, the probability that at least one has not failed its self-attestation is given by 
\[
    1 - \prod_{k=1}^{\zeta_d^\ell} \Pr[D_k \mid \Gamma].
\]
In the worst case, where every malicious ECU manipulates the challenge value, this corresponds to the probability of a false positive.

\noindent \textit{For false negatives}, when a compromised node is challenged and at least one verifier is benign, the round reports the node as valid with probability $1 - Pr[D \mid \Gamma]$: given that the challenged node contains malware, \texttt{TApp} will not reveal the value $C$ to its associated \texttt{UApp} if any self-attestation has produced invalid results, with probability $Pr[D \mid \Gamma]$. Given that there is a benign verifier, a false negative occurs only when self-attestation misses the malware, so its probability is $1 - Pr[D \mid \Gamma]$. Note that such a round does not prevent later detection, as the node remains subject to future self and decentralized attestations.

\noindent \textit{For eventual recovery}, we have established that a node failing local self-attestation is detected when it is selected as the challenged node and a selected verifier is benign. In the worst case, all in-vehicle computers are compromised. By Assumption 3, the vehicle operator can reboot \texttt{MC}. After being restored, \texttt{MC} verifies and repairs the zone controllers. Then, the benign zone controllers will verify ECUs and help repair them.

\subsection{Discussion}

\noindent\textbf{Context-aware reboot policy}. 
While coordinated repair provides the mechanism to initiate a reboot, the logic for when \texttt{MC} should issue that signal remains an open challenge. In Level 4 or 5 autonomous vehicles, this decision may mostly be automated, so that the vehicle smartly schedules reboot or drives into a safe state before rebooting the target node. In lower levels of automation, \texttt{MC} might simply notify the driver in all cases, so that they can pull over before manually issuing a reboot command. Further, the reboot timing depends heavily on the specific ECU being repaired~\cite{andreasson2022device}. While non-critical ECUs may reboot during runtime with low risk, safety-critical ECUs require a more complex decision. Importantly, by decoupling the repair mechanism from the reboot policy, our design allows manufacturers to implement various strategies.

\noindent\textbf{An application to VANET security}. 
Vehicular ad hoc networks (VANETs) enable V2V and V2I communications, but malicious code injection can cause compromised vehicles to mislead not only the impacted vehicle but also the entire VANET. Previous VANET security solutions~\cite{oham2021bferl, kaster2024automotive,hellemans2025spark} propose computing a signature on the entire software on each vehicle computer or blockchain based consensus mechanisms. However, these strategies are expensive and can violate the real-time requirements of the vehicle. A slight modification of~\ours{} can efficiently enable secure participation in VANETs. Specifically, an external entity may simply perform remote-attestation with \texttt{TApp} in \texttt{MC}, and obtain, through a secure channel, a view of the entire vehicle software integrity. The layered approach of our solution becomes analogous to the merkle tree used by other approaches~\cite{oham2021bferl,kaster2024automotive}.

\noindent\textbf{Challenging multiple nodes}. 
To significantly improve repair time, we introduce $a_\ell$, which is the number of nodes challenged in layer $\ell$ during each decentralized-attestation epoch. This requires two small modifications of~\ours{}: 1) Algorithm~\ref{alg:challenge-generation} generates $a_\ell$ challenged node IDs and their respective challenge responses. 2) Each verifier validates all $a_\ell$ challenged nodes. 
The significant impact of $a_\ell$ on the mean time to repair is shown in Table~\ref{tab:layer1-sweep}.

%% file: Source/6-evaluation.tex
\section{Implementation and Evaluation}
\label{sec:eval}
\subsection{Implementation}

\begin{figure}[tb]
  \centering

  \resizebox{0.80\linewidth}{!}{%
    \begin{minipage}{\linewidth}
      \centering

      \begin{minipage}[c]{0.58\linewidth}
        \centering
        \includegraphics[width=\linewidth]{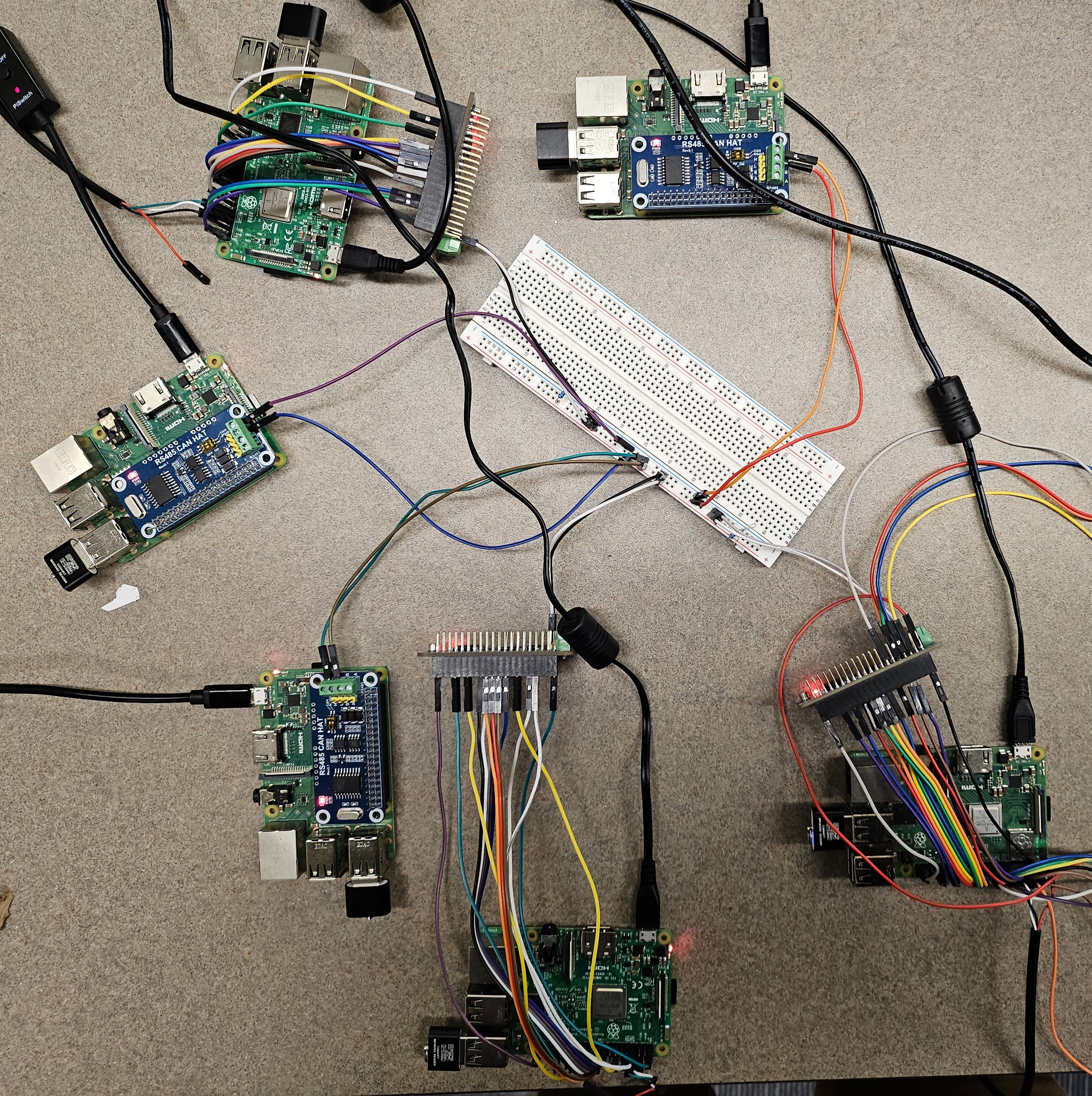}
      \end{minipage}
      \hfill
      \begin{minipage}[c]{0.39\linewidth}
        \centering
        \includegraphics[height=1.5\linewidth,keepaspectratio]{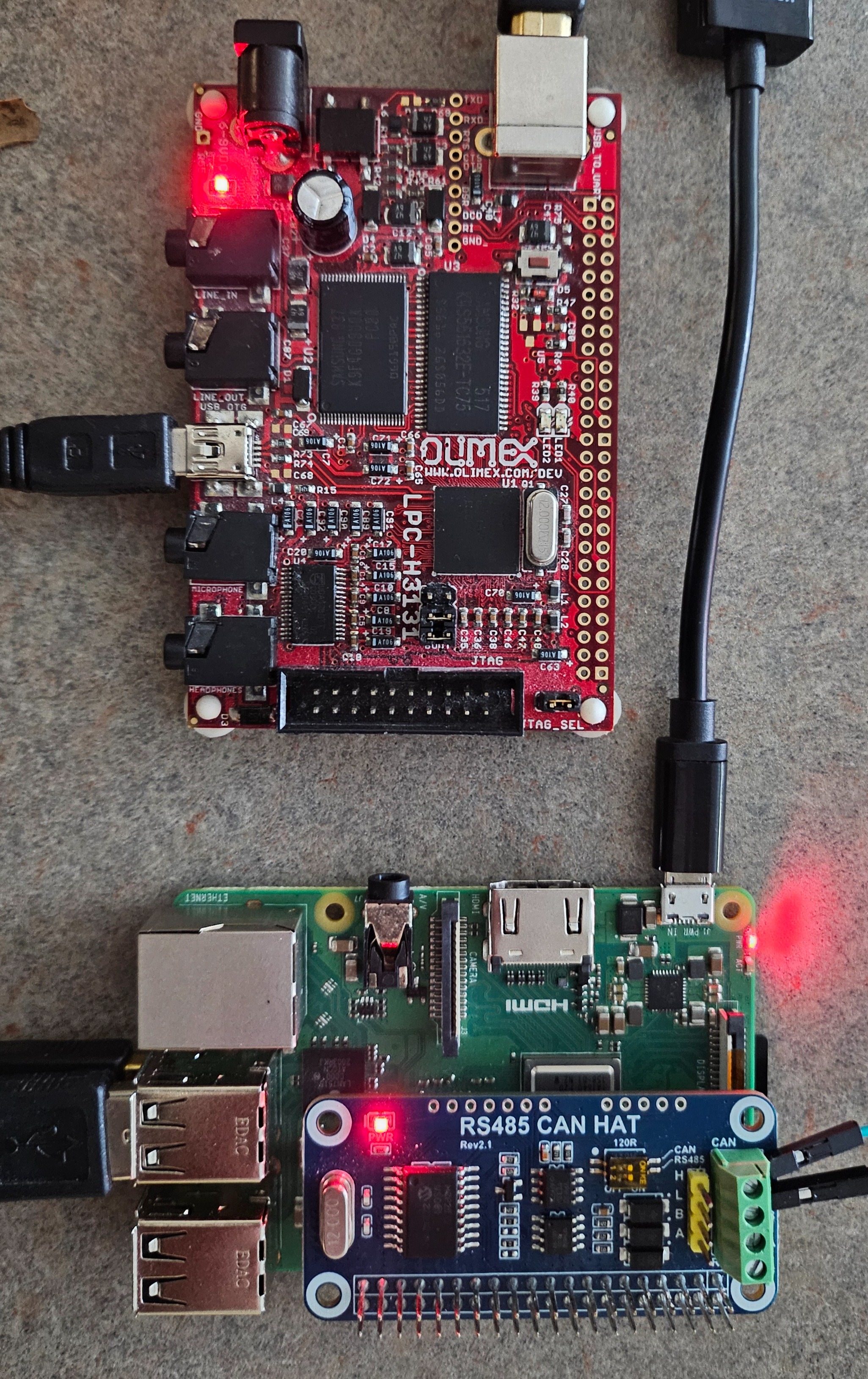}
      \end{minipage}
    \end{minipage}%
  }

  \caption{Left: the 6-node Raspberry Pi 3/3B+ testbed connected through a CAN bus. Right: A Raspberry Pi attached to an LPC-H3131 development board used as flash storage.}
  \label{fig:testbed}
\end{figure}
To demonstrate the feasibility of our design, we have implemented \ours{} on real-world computing hardware. In our implementation, we simulate 6 ECUs using Raspberry Pi, where four ECUs are simulated using Pi 3B+, and two are Pi 3B (with a 1.2GHz (3B+ is 1.4GHz) 64-bit quad-core ARM Cortex-A53 CPU, and 1GB LPDDR2 SDRAM) with TrustZone enabled. Communication between nodes is implemented using the RS485 CAN hat, which supports CAN 2.0. We implemented \texttt{UApp} under our modification of the standard OP-TEE (Open Portable Trusted Execution Environment) OS installation. We implemented \texttt{TApp} in the ARM TrustZone secure world. For our self-attestation and self-repair schemes, we used the constructions from~\cite{dafoe2025hardware}. \texttt{FMC} was implemented on a USB 2.0 header development board LPC-H3131 (with ARM9 32-bit RM 926EJ-S, 180Mhz, 32MB of SDRAM and 512MB of NAND flash). We have ported the OpenNFM open source NAND flash manager to LPC-H3131 and modified it for our FMC implementation.

\subsection{Experimental Evaluation}

In addition to implementing \ours{}, we have performed an experimental analysis to evaluate the cost. For self-attestation, our experimental results are summarized in Table~\ref{tab:self-attestation-experiment}. There is a setup phase, which consists of first verifying a signature over the entire firmware image, then, if this is valid, generating verification tags over each block. Immediately, we notice that this is the most expensive operation in our design, taking up to $5.720$ seconds to initialize a $400\text{KiB}$ firmware image. This high cost is expected due to the use of expensive cryptographic operations performed in the TrustZone. This high cost justifies our design choice to run this operation in the parent node, where far more computational power is available than a basic ECU. Additionally, we note that this operation only occurs once and therefore will not impact normal vehicle behavior. In contrast, the regular self-attestation operations (i.e. Challenge generation, proof generation, and proof verification) are very quick due to the spot-checking technique used by~\cite{dafoe2025hardware}, allowing runtime self-attestation to meet the real-time in-vehicle requirement. Additionally, we see that the costs of the attestation-phase scale roughly linearly with the size of the data challenged.

\begin{table}[t]
\vspace*{0.05in}
\centering
\setlength{\tabcolsep}{7pt}
\resizebox{.85\columnwidth}{!}{
\begin{tabular}{|c|c|c|c|}
\hline
Procedure & Component & Size & Time (ms) \\ \hline
\multirow{2}{*}{Setup (one time)} & \multirow{2}{*}{Parent Node \texttt{TApp}} & 200KiB & 3,996 \\ \cline{3-4}
 & & 400KiB & 5,720 \\ \hline
\multirow{2}{*}{Challenge Generation} & \multirow{2}{*}{\texttt{TApp}} & 100KiB & 17.9 \\ \cline{3-4}
 & & 200KiB & 21.2 \\ \hline
\multirow{2}{*}{Proof Generation} & \multirow{2}{*}{\texttt{UApp}} & 100KiB & 19.2 \\ \cline{3-4}
 & & 200KiB & 39.9 \\ \hline
\multirow{2}{*}{Proof Verification} & \multirow{2}{*}{\texttt{TApp}} & 100KiB & 23.8 \\ \cline{3-4}
 & & 200KiB & 31.6 \\ \hline
\end{tabular}}
\caption{Runtime for the Self-Attestation phases with different sizes. For the Setup phase, the size considered is the full data size. For the Attestation phase, (including challenge generation, proof generation, and proof verification) the size considered is the size of the challenged data (i.e. $c \cdot |b_i|$).}
\label{tab:self-attestation-experiment}
\end{table}

For self-repair, our experimental results are summarized in Table~\ref{tab:self-repair-experiment}. There are two dedicated operations for self-repair: Notification and Rollback, both implemented in \texttt{FMC}. The notification consists of simply reading a value stored at a special address, generating a corresponding value, and checking whether they match. This is a very simple operation and so has the lowest cost among the operations in our design. Importantly, this is a periodic operation, and so the negligible cost demonstrates that the notification will not interfere with regular storage operations. In contrast, the rollback operation is far more expensive, but occurs only when malware is present, in order to restore the latest benign firmware. However, only the mappings need to be reverted to restore the entire data. 

\begin{table}[t]
\centering
\setlength{\tabcolsep}{7pt}
\resizebox{.85\columnwidth}{!}{
\begin{tabular}{|c|c|c|c|}
\hline
Procedure & Component & Time (ms) \\ \hline
Notification & \texttt{FMC} & 2.3\\ \hline
Rollback (Mapping Restoration) & \texttt{FMC} & 1,315.5\\ \hline
\end{tabular}}
\caption{Run time for the Self-Repair phases. The rollback operation restores the entire storage available in LPC-H3131. Note that the throughput is limited by the old USB 2.0 interface used between the LPC-H3131 and Raspberry Pi.}
\label{tab:self-repair-experiment}
\end{table}

For decentralized attestation, our experimental results are summarized in Table~\ref{tab:decentralized-attestation-expirement}. Our immediate observation is that these operations are very fast and incur a very small cost on top of the self-attestation, while increasing the overall security dramatically. Challenge generation and response generation both consist of simply generating random values from shared secrets. However, challenge generation is run on \textit{all} nodes, while response generation is performed only by the challenged node and selected verifiers (challenge generation simply determines the role of each node). Given this fact, and the marginally higher runtime of challenge generation, it is by far the most expensive operation in decentralized attestation. However, since we observe that it only takes $14.5\text{ms}$, the cost is still quite small. 

\begin{table}[t]
\vspace*{0.05in}
\centering
\setlength{\tabcolsep}{7pt}
\resizebox{.85\columnwidth}{!}{
\begin{tabular}{|c|c|c|}
\hline
Procedure & Component & Time (ms) \\ \hline
Challenge Generation & \texttt{TApp} & 14.5 \\ \hline
Response Generation & \texttt{TApp} & 7.7 \\ \hline
Proof Verification & \texttt{TApp} & 7.8 \\ \hline
\end{tabular}}
\caption{Run time for the Decentralized-Attestation phases. The time for challenge generation depends on $|D_d^\ell|$ and $v_d^\ell$. The time for response generation and proof verification depend only on the security parameter $l$.}
\label{tab:decentralized-attestation-expirement}
\vspace{-5pt}
\end{table}

\begin{table}[t]
  \centering
  \begin{tabular}{|l|r|r|r|r|}
    \hline
    System & RW & RR & SW & SR \\
    \hline
    OpenNFM & 1723 & 2029 & 2223 & 2763 \\
    \hline
    \ours{} & 1644 & 2130 & 2221 & 2779 \\
    \hline
  \end{tabular}
  \caption{Throughput comparison between $\ours$ and original OpenNFM. Throughput is reported in KiB/s.}
  \label{tab:compareThroughput}
  \vspace{-15pt}
\end{table}

\noindent \textbf{Performance impact on the flash storage.} In addition to implementing notification as a regular operation on \texttt{FMC}, we modified the garbage collection policy, and explicitly maintain a backup mapping table. To evaluate the impact of these changes on regular storage operations, we have measured the throughput of our modified version of OpenNFM with the original version. These results are summarized in Table~\ref{tab:compareThroughput}. We observe that our design largely maintains baseline performance. The RW throughput is -4.6\% relative to the original, while RR improves by +5.0\%, and SW and SR are effectively unchanged. These differences are small ($\leq5\%$) and may fall within typical run-to-run variability; additionally, a slight write speed reduction is expected due to the extra work to maintain the backup mapping table.

%% file: Source/8-related.tex
\section{Related Work}
\label{sec:related}

\subsection{In-vehicle ECU Firmware Attestation}
A few existing works use secure boot in automotive ECUs~\cite{gui2018hardware,sanwald2019secure,anand2024achieving}, checking the firmware image before boot. In particular, Gui et al.~\cite{gui2018hardware} propose using TPM to employ a core root of trust measurement, which computes HMAC digests on the firmware image at boot and compares them with a stored golden measurement. However, secure boot requires computing a signature on the entire firmware image, which is not scalable and generally cannot meet the $50-100$ms boot time requirement~\cite{kim2017secure,ep3923168b1}. To mitigate this, Kim et al.~\cite{kim2017secure} propose moving the application memory verification to the background (in an HSM) with prioritized blocks, allowing the ECUs to become operational before completing the complete software verification. Rogala~\cite{ep3923168b1} proposes performing secure boot at shutdown, essentially pre-verifying the firmware so that the strict boot-time requirement can be met. Although these solutions try to mitigate the cost associated with secure boot, they still suffer from checking the entire firmware image. 
Oguma et al.~\cite{oguma2008new} propose using a dedicated master ECU to perform internal remote-attestation of other ECU's firmware at vehicle startup. Similarly, Kohnh\"auser et al.~\cite{kohnhauser2019ensuring} rely on a master ECU to verify the firmware of other ECU's. Their design requires verification of a pre-defined set of safety-critical ECUs before ignition is allowed.

All the aforementioned designs suffer from a few common limitations: 1) they suffer from a single point of failure; and 2) they do not consider the hierarchical nature of the in-vehicle networks; and 3) they suffer from high computational overhead, since each ECU still needs to compute a hash over the entire firmware image; and 4) they do not integrate a repair component. SPARK~\cite{hellemans2025spark} introduces a hierarchical swarm attestation that focuses on privacy preservation and allows external attestation for participation in VANETs. It can address the first and second limitations mentioned above, but it still suffers from the high computational cost resulting from hashing the entire firmware on each ECU. Other works~\cite{kaster2021sliced, nasser2019accelerated} introduce various probabilistic self-attestation to meet the real-time requirement, but are still vulnerable to a single point of failure and do not integrate a repair component. 

Khodari et al.~\cite{khodari2019decentralized} introduced a decentralized firmware attestation design, in which each ECU can challenge any other ECU, as long as the valid signature (among other metadata) is stored in its local database. The challenged ECU will then generate a signature on the firmware image (using some trusted hardware extension) and transmit it to the verifier. Compared with $\ours$, their design suffers from a few limitations: 1) verifier storage grows linearly with the number of ECUs it may challenge; and 2) attestation requires computing a signature over the entire firmware image, which is expensive; and 3) the protocol involves multiple communication rounds rather than a single lightweight message; and 4) the design assumes a flat network topology and does not exploit the hierarchical structure of in-vehicle networks; and 5) they do not incorporate a repair component.

\subsection{In-vehicle ECU Firmware Repair}

To restore the system after malware attacks, Mansor et al.~\cite{mansor2015dont} proposed temporarily backing up the previous firmware during updates, and, if the installation-time checks fail, the old firmware is restored. However, this solution only supports recovery immediately after flashing and cannot address runtime compromise that occurs long after a successful update. 
Kwon et al.~\cite{kwon2018mitigation} proposed integrating an intrusion detection system and moving compromised ECUs into a ``safe mode'' before rebooting. Dafoe et al.~\cite{dafoe2024enabling} proposed a self-repair mechanism for ECUs
together with an intrusion detection node that can notify a trusted application in compromised ECUs to orchestrate rollback. They further improved the design by incorporating a probabilistic self-attestation in their subsequent work~\cite{dafoe2025hardware}. 
However, all these approaches suffer from three limitations: 1) they are vulnerable to a single point of failure, and 2) they do not explicitly verify the entire state of the vehicle and instead provide only local integrity checks, and 3) they do not support context-aware rebooting decisions, because the ECU status is not propagated to the rest of the vehicle. 

%% file: Source/appendix2.tex
\section{Extended Security analysis of $\ours$}

\label{sec:appendix:security}


\subsection{Decentralized-attestation}
\label{sec:security:decentralized}

In each domain $D_d^\ell$ (where $1 \leq \ell \leq 2$ and $1 \leq d \leq n_\ell$, with $n_\ell$ denoting the number of domains in layer $\ell$), a periodic decentralized-attestation will take place, during which one node in the domain is challenged, and $v_d^\ell$ nodes act as verifiers. The challenged node should respond with the specific value $C$ if all self-attestations have returned valid results since the last reboot. For this mechanism to be secure, we have three security requirements, which we analyze as follows: 

\medskip
\noindent\textbf{Property 1 (Bounded false positives).} Let $\zeta_d^\ell$ be the number of malicious verifiers in domain $D_d^\ell$ ($\zeta_d^\ell \leq v_d^\ell$), indexed $k = 1, \dots, \zeta_d^\ell$, and let $\Pr[D_k \mid \Gamma]$ be the probability that corrupted firmware is detected on verifier $k$. The probability of a false positive is at most
\[
    1 - \prod_{k=1}^{\zeta_d^\ell} \Pr[D_k \mid \Gamma].
\]

\noindent\textit{Proof sketch.}
Recall that $\Pr[D \mid \Gamma]$ is the probability that corrupted firmware is detected on a given node. A false positive can only occur when at least one verifier returns an invalid attestation report. This occurs in one of three mutually exclusive cases:

\smallskip
\noindent\textit{(a) Challenged node withholds $C$.} \texttt{TApp} in the challenged node does not reveal $C$ to \texttt{UApp} (i.e., $C' \neq C$). Self-attestation necessarily failed, so malware is definitely present. This is a true positive unless \texttt{UApp} independently guesses $C$, which succeeds with only negligible probability $\frac{1}{2^\kappa}$.

\smallskip
\noindent\textit{(b) Challenged node's \texttt{UApp} tampers with $C$.} \texttt{TApp} reveals $C$ (having detected no malware), but a compromised \texttt{UApp} modifies or drops it in transit (i.e., $C' \neq C$). The challenged node contains malware, so this is again a true positive.

\smallskip
\noindent\textit{(c) Verifier's \texttt{UApp} tampers with $C'$.} The challenged \texttt{UApp} is benign ($C' = C$), but malware in verifier $k$ causes its \texttt{UApp} to modify or drop $C'$ in transit (i.e., $C'' \neq C$). This is the only source of false positives. The malware in verifier $k$ is independently detected via spot-checking with probability $\Pr[D_k \mid \Gamma]$, in which case \texttt{TApp} will not generate an invalid report. Otherwise (probability $1 - \Pr[D_k \mid \Gamma]$), the compromised verifier may produce a false invalid attestation report.

\smallskip
Since spot-checks depend on the local state of each node, the detection events are independent. The probability that all $\zeta_d^\ell$ malicious verifiers are detected is then $\prod_{k}^{\zeta_d^\ell} \Pr[D_k \mid \Gamma]$, and the probability that \emph{at least one} remains \textit{undetected} and causes a false positive is at most
\[
    1 - \prod_{k=1}^{\zeta_d^\ell} \Pr[D_k \mid \Gamma].
\]
This is an upper bound, since malicious nodes will not necessarily perform such an attack.

\smallskip
Note that even in the rare case of a false positive, the system response is to reboot the challenged node at a ``safe'' time. This has low impact compared to missing an actual compromise, and the probability of such events is low by the above bound.

\medskip
\noindent\textbf{Property 2 (Decentralized attestation introduces no false negatives).} If a challenged node has failed self-attestation and at least one verifier is benign, then the compromise is detected except with negligible probability

\smallskip
\noindent\textit{Proof sketch.} Suppose a node has been challenged, and its self-attestation has failed. As established in Case (a), TApp in the challenged node will not reveal $C$ to UApp. Additionally, $C$ cannot be forged with probability higher than $1/2^\kappa$. Therefore, TApp in a benign verifier will not receive $C$ (aside from a negligible probability) and will therefore produce an invalid attestation report.

\medskip
\noindent\textbf{Property 3 (Per-round false-negative rate).} Suppose a malicious node is selected as the challenged node in a given decentralized-attestation round. Then that round reports the node as valid with probability at most $1 - \Pr[D \mid \Gamma]$.

\smallskip
\noindent\textit{Proof sketch.} There are two cases, depending on whether self-attestation has already failed on the challenged node:

\smallskip
\noindent\textit{(a) Self-attestation has failed} (probability $\Pr[D \mid \Gamma]$). By Property 2, the benign verifier detects the compromise. This case contributes no false negatives.

\smallskip
\noindent\textit{(b) Self-attestation has not detected the compromise} (probability $1 - \Pr[D_c \mid \Gamma]$). \texttt{TApp} perceives no invalid result and honestly reveals $C$. If the compromised \texttt{UApp} forwards $C$ unmodified, the verifiers receive the correct value and report the node as valid, resulting in a false negative. If instead it modifies or drops $C$, the round reports the node as invalid (Property 1, case (b)) and the compromise is caught.

\smallskip
Since a false negative occurs only in case (b), the per-round false-negative probability is exactly $1 - \Pr[D \mid \Gamma]$.

\smallskip 
Recall that $\Gamma$ considers the number of self-attestations performed since the compromise. This can be 0, in which case $1 - \Pr[D \mid \Gamma] = 1 - 0 = 1$. Additionally, A false-negative round does not prevent later detection: the node remains subject to self and decentralized attestations and will be challenged again. $\Pr[D \mid \Gamma]$ increases with each subsequent attestation and the probability of indefinite non-detection goes to zero. The mean time, beginning from node compromise, during which a decentralized-attestation challenge results in a false negative is quantified by $W_1$ in \S~\ref{sec:security:repair}.

\subsection{Coordinated-repair}
\label{sec:security:repair}

For coordinated-repair to be secure, we want to ensure that over time, any node that is definitely compromised (i.e., it has failed self-attestation or \texttt{TApp} is disabled) is eventually rebooted into a benign state. To do this, the invalid attestation report generated by a verifier node needs to first successfully reach \texttt{MC}. Since \texttt{MC} can initiate repair of any node (by Assumption 2), this is sufficient for repair to take place. We provide a proof sketch below (by strong induction on the layers~\footnote{Note that in this section, we use a generalized system model, with a vehicle architecture consisting of $n$ layers.}):

\medskip
For the following property, let $c^\ell_k$ denote the $k$th in-vehicle computer in layer $\ell$, where layer $n$ contains only \texttt{MC}. The argument assumes that once a node is repaired, it remains benign long enough to participate in the repair of nodes in lower layers.

\noindent\textbf{Property 4 (Eventual repair).}
For any in-vehicle computer $c^\ell_k$, if $c^\ell_k$ is definitely compromised (i.e., it has failed self-attestation or \texttt{TApp} is disabled), it will be rebooted via coordinated repair within finite time.

\smallskip
\noindent\textit{Proof sketch (strong induction on layers).} For the base case, layer $n$ consists solely of \texttt{MC}. If \texttt{MC}'s firmware is compromised, then \texttt{TApp} in \texttt{MC} detects this via self-attestation eventually (since $\Pr[D \mid \Gamma]$ asymptotically approaches $1$ as the number of self-attestations increases). Then by Assumption 3, \texttt{TApp} can alert the vehicle operator, who can move the vehicle into a safe state (e.g., parked) and manually reboot \texttt{MC}.

\smallskip
For the inductive step, suppose that for some $i$ with $1 < i \leq n$, any detected malware in layers $n$ through $i$ has been repaired via self-repair and coordinated repair (inductive hypothesis). Suppose some node $c_k^{i-1}$ in layer $i-1$ is compromised and has failed self-attestation. Since all nodes in layers $n$ through $i$ are eventually repaired, the parent of $c_k^{i-1}$ is eventually repaired. After this point, either a benign node in layer $i-1$ or the parent itself will be selected to verify $c_k^{i-1}$ during decentralized attestation. By Property 1--2 (\S\ref{sec:security:decentralized}), this will be detected by the benign verifiers, and an invalid attestation report will be propagated upward. Since all compromised nodes in upper layers have been repaired (inductive hypothesis), this message is guaranteed to reach \texttt{MC} intact. \texttt{MC} can then determine a safe window in which to reboot $c_k^{i-1}$, which is feasible by Assumption 2.

\smallskip
\noindent\textit{Remark.}
This proof sketch implicitly assumes that once a node is repaired, it remains benign long enough to facilitate repair of lower layers.

While we can ensure that repair will occur eventually, we also need to ensure that this repair is timely. To determine this, we want to compute the mean time to repair for each layer ($MTTR_\ell$). We define the mean time to repair as the time until repair is feasible. That is, the time until \texttt{TApp} in \texttt{MC} is aware of malware on a given node. At this point, repair may take place at any point, decided by \texttt{MC}.

First, we note that for a given compromised node in layer $i$, there are a few conditions to guarantee that an invalid attestation report will reach \texttt{MC}: (1) The node has failed a self-attestation, (2) The node is selected to be challenged during decentralized attestation, (3) a benign verifier is selected during this challenge, (4) a specific node in each layer $k$ where $i < k \leq n$ is benign (i.e. the parent controllers up to \texttt{MC}). Note that (3) and (4) are required because we assume the worst case attacker, which performs a DoS attack on broadcasting an invalid attestation report. To compute $MTTR_\ell$, we want to determine the mean time until all of these conditions have been met. We compute $MTTR_\ell$ in two different ways: analytically by solving the fixed point presented below, and via a Monte Carlo simulation. The two results agree, as shown in Tables~\ref{tab:baseline-layers} and~\ref{tab:layer1-sweep}. Both models consider a more general design than we presented in the main text. Specifically, we consider challenging $a_\ell$ nodes per decentralized-attestation epoch (\S~\ref{sec:discussion}). Additionally, we assume that rebooting takes place at the instant a benign $\texttt{MC}$ receives a invalid decentralized-attestation report. 

\noindent\textbf{Analytic solution.} Let $P_m = \Pr[D\mid\Gamma]$. First, (for condition (1)) we model the number of self-attestations until corruption is detected locally as a random variable ($N_{SA}$) that follows a geometric distribution. The \textit{expected value} for the number of self-attestations until corruption is detected locally is therefore 
$$
\Large \mathbb{E}[N_{SA}] = \sum_{k = 1}^{\infty}\big[(P_m\cdot(1-P_m)^{k-1})\cdot k \big] = \frac{1}{P_m}.
$$ 

To deduce the mean \textit{time} until self-attestation detects malware, we model the time until corruption is detected locally as a random variable $\tau_{SA}$, where $\tau_{SA} = T_{SA} \cdot N_{SA}$. The expected value is $$
\displaystyle \mathbb{E}[\tau_{SA}] = \mathbb{E}[N_{SA}] \cdot T_{SA} = \frac{T_{SA}}{P_m}.
$$ 
We simply take the reciprocal to get the rate (per unit time) of detection by self-attestation: 
$$
\lambda_1 = \frac{1}{\mathbb{E}[\tau_{SA}]} =\frac{P_m}{T_{SA}}.
$$

Since malware installation occurs at a time independent of the self-attestation schedule, the first check after installation does not land a full $T_{SA}$ later on average, but only $T_{SA}/2$ later. We therefore define $W_1$, the expected time until self-attestation first detects malware, as
$$
W_1 = \frac{1}{\lambda_1} - \frac{T_{SA}}{2}.
$$

In addition (for condition (2)), we model detection via decentralized attestation. We model the number of decentralized-attestation rounds until the malicious node is challenged in domain $D_d^\ell$ by the random variable $N_{DA}^{(d,\ell)}$ following a geometric distribution. The expected value for the number of decentralized attestations until a given compromised node is challenged is $\mathbb{E}[N_{DA}^{(d,\ell)}] = |D_d^\ell|$. Let $\tau_{DA}^{(d,\ell)}$ be a random variable representing the time until the compromised node is challenged, so that $\tau_{DA}^{(d,\ell)} = T_\ell\cdot N_{DA}^{(d,\ell)}$. The expected time until being challenged is therefore $$
\mathbb{E}[\tau_{DA}^{(d,\ell)}] = T_\ell \cdot \mathbb{E}[N_{DA}^{(d,\ell)}] = T_\ell \cdot |D_d^\ell|.
$$
Thus, the rate of detection by decentralized attestation is $$\lambda_2 = \frac{a_\ell}{\mathbb{E}[\tau_{DA}^{(d,\ell)}]}=  \frac{a_\ell}{T_\ell \cdot |D_d^\ell|.}$$ 

Decentralized-attestation challenges occur independently of when self-attestation detects malware, so we define $W_2$, the expected time until the first decentralized-attestation challenge after detection, as
$$
W_2 = \frac{1}{\lambda_2} - \frac{T_\ell}{2}.
$$

Then, the mean time until both conditions (1) and (2) are satisfied is then
simply
$$
W_1 + W_2.
$$

For condition (3), we also need a benign node to be selected as a verifier. To do this we take into account the availability of the nodes in the same layer:
$$
A_\ell = \frac{MTTF_\ell}{MTTF_\ell + MTTR_\ell},
$$
where $MTTF_\ell$ is the mean time to failure for a given node in layer $\ell$. Given this, the probability, at any time, that at least one verifier is benign is
$$
P_s = 1 - (1 - A_\ell)^{v_d^\ell}.
$$
By Property 2, if a compromised node is challenged and at least one verifier is benign, the malware is guaranteed to be detected. Thus $P_s$ is also the probability that a given challenge round successfully detects the malware, and $1-P_s$ is the probability that a challenge round fails to detect it (i.e., every verifier selected for that round happens to be malicious). If a challenge round fails, detection must wait for a subsequent decentralized-attestation challenge, which arrives after a further expected time $1/\lambda_2$; this may again fail with probability $1-P_s$, and so on. Accounting for this retry process, the expected additional time, after self-attestation detects malware, until conditions (2) and (3) are satisfied is
$$
\begin{aligned}
&W_2 + (1-P_s)\frac{1}{\lambda_2} + (1-P_s)^2\frac{1}{\lambda_2} + \cdots \\
&\quad = W_2 + \frac{1}{\lambda_2}\sum_{k=1}^{\infty}(1-P_s)^k \\
&\quad = W_2 + \frac{1}{\lambda_2}\cdot\frac{1-P_s}{P_s}.
\end{aligned}
$$
So that the expected time until conditions (1), (2), and (3) are met is
$$
W_1 + W_2 + \frac{1-P_s}{\lambda_2 \cdot P_s}.
$$

For condition (4), we also need the parent controller in the layers $\ell+1$ through $n$ to be benign for the message to be received by MC. I.e. a particular node in each of the higher layers must be benign. This depends on the value of $A_k$ for $\ell + 1 \leq k \leq n$. In fact, the probability that all of the parents up to the main controller are benign at any given time is
$$
\prod_{k=\ell+1}^n A_k.
$$
To incorporate this, we can simply modify $P_s$:
$$
P_s = \big[1 - (1-A_\ell)^{v_d^\ell}\big] \cdot \prod_{k=\ell+1}^n A_k.
$$
Therefore, we have, as the expected time until conditions (1), (2), (3), and (4) are all met:
$$
MTTR_\ell = W_1 + W_2 + \frac{1-P_s}{\lambda_2 P_s}.
$$
However, there is a dependency of $MTTR_\ell$ on $A_\ell$, and vice versa. But, we have
\begin{align*}
A_\ell &= \frac{MTTF_\ell}{MTTF_\ell + MTTR_\ell} \\[0.5em]
&\implies (MTTF_\ell + MTTR_\ell)A_\ell = MTTF_\ell \\[0.5em]
&\implies A_\ell \cdot MTTF_\ell + A_\ell \cdot MTTR_\ell = MTTF_\ell \\[0.5em]
&\implies A_\ell \cdot MTTR_\ell = MTTF_\ell - A_\ell \cdot MTTF_\ell \\[0.5em]
&\implies A_\ell \cdot MTTR_\ell = MTTF_\ell(1 - A_\ell) \\[0.5em]
&\implies MTTR_\ell = MTTF_\ell \cdot \frac{1-A_\ell}{A_\ell},
\end{align*}
so that by combining the two results for $MTTR_\ell$, we have
$$
MTTF_\ell \cdot \left(\frac{1-A_\ell}{A_\ell}\right) = W_1 + W_2 + \frac{1-P_s}{\lambda_2 \cdot P_s},
$$
which implies that
$$
MTTF_\ell \cdot \left(\frac{1-A_\ell}{A_\ell}\right) - W_1 - W_2 - \frac{1-P_s}{\lambda_2 \cdot P_s} = 0.
$$
To solve for $A_\ell$, we can use a fixed point approach (e.g. Newton's method). Using this, we can compute $MTTR_\ell$. We note that to compute $MTTR_\ell$, we first need to compute $MTTR_k$ for $\ell < k \leq n$, since it is defined recursively. For the base case, since repair in MC depends only on self-attestation, we have
$$
MTTR_n = W_1.
$$

\noindent\textbf{Monte Carlo Simulation.} To validate the analytical results, we implemented a Monte Carlo simulation of node compromise, self-attestation, decentralized-attestation, and coordinated-repair. Each simulation run samples compromise times, self-attestation results, decentralized-attestation challenges, verifier availability, and parent controller availability. A compromised node was only repaired if all four conditions above were simultaneously met. We tracked the time of compromise and time of repair for each instance, and computed $MTTR_\ell$ as the empirical mean time from node compromise until the compromised node is reported to \texttt{MC}.

Table~\ref{tab:baseline-layers} shows the baseline parameters used when computing $MTTR_\ell$ using both the analytical results and Monte Carlo Simulation. The baseline results for all layers are shown in Table~\ref{tab:baseline-layers}. We also vary several key parameters and report the corresponding value of $MTTR_1$ in Table~\ref{tab:layer1-sweep}. These results demonstrate both the timely nature of our design and the close agreement between the analytical and Monte Carlo estimates. Across the baseline layer comparison, the maximum relative error is $0.35\%$; across the $MTTR_1$ parameter sweep, the maximum relative error is $0.49\%$.

  \begin{table}[tb]
      \centering
      \begin{tabular}{lll}
      \toprule
      Parameter & Value & Description \\[0.35em]
      \midrule
      $n$ & $3$ & Number of hierarchy layers \\[0.35em]
      $\Pr[D \mid \Gamma]$ & $0.982$ & Detection probability per attestation \\[0.35em]
      $T_{SA}$ & $4\,\mathrm{s}$ & Self-attestation period \\[0.35em]
      $MTTF_\ell$ & $5400\,\mathrm{s}$ & Mean time to failure for each layer \\[0.35em]
      $T_1, T_2$ & $2\,\mathrm{s}$ & Decentralized-attestation periods \\[0.35em]
      $T_3$ & $T_{SA}$ & Top-layer attestation period \\[0.35em]
      $|D_d^2|$ & $4$ & Number of zone controllers \\[0.35em]
      $|D_d^1|$ & $20$ & Number of ECUs in each domain \\[0.35em]
      $v_d^2$ & $2$ & Verifiers for layer 2 \\[0.35em]
      $v_d^1$ & $5$ & Verifiers for layer 1 \\
      \bottomrule
      \end{tabular}
      \caption{Baseline parameters used for the computations.}
      \label{tab:baseline-parameters}
  \end{table}

 \begin{table}[tb]
      \centering
      \begin{tabular}{llrrr}
      \toprule
      Param. & Value & Analytic & MC mean & Difference \\
      \midrule
      $\mathrm{MTTF}$ & 5400 & 41.16 & 41.15 & 0.01\% \\
       & 2000 & 41.30 & 41.32 & 0.05\% \\
       & 800 & 41.63 & 41.71 & 0.17\% \\
       & 300 & 42.58 & 42.78 & 0.49\% \\
       & 100 & 45.84 & 46.06 & 0.48\% \\
      \midrule
      $a_\ell$ & 1 & 41.16 & 41.15 & 0.01\% \\
       & 2 & 21.10 & 21.11 & 0.07\% \\
       & 4 & 11.08 & 11.08 & 0.00\% \\
       & 8 & 6.08 & 6.08 & 0.05\% \\
      \midrule
      $|D_d^1|$ & 10 & 21.11 & 21.13 & 0.06\% \\
       & 20 & 41.16 & 41.15 & 0.01\% \\
       & 40 & 81.24 & 81.26 & 0.03\% \\
       & 80 & 161.40 & 161.41 & 0.01\% \\
      \midrule
      $v_1$ & 1 & 41.46 & 41.51 & 0.12\% \\
       & 2 & 41.16 & 41.15 & 0.01\% \\
       & 5 & 41.16 & 41.15 & 0.01\% \\
       & 10 & 41.16 & 41.18 & 0.05\% \\
      \bottomrule
      \end{tabular}
      \caption{Layer-1 MTTR under parameter sweeps. MC means are over 50 random seeds, each simulating $10^7\,\mathrm{s}$.}
      \label{tab:layer1-sweep}
  \end{table}

  \begin{table}[tb]
      \centering
      \begin{tabular}{lrrr}
      \toprule
      Layer & Analytic & MC mean & Difference \\
      \midrule
      Layer 1 & 41.16 & 41.15 & 0.01\% \\
      Layer 2 & 9.08 & 9.04 & 0.35\% \\
      Layer 3 & 2.07 & 2.08 & 0.16\% \\
      \bottomrule
      \end{tabular}
      \caption{Baseline analytical and Monte Carlo MTTR across all layers. MC means are over 50 random seeds, each
      simulating $10^7\,\mathrm{s}$.}
      \label{tab:baseline-layers}
  \end{table}

%% file: refs.bib
@inproceedings{antinyan2020revealing,
  author    = {Vard Antinyan},
  title     = {Revealing the Complexity of Automotive Software},
  booktitle = {Proceedings of the 28th {ACM} Joint European Software Engineering Conference and Symposium on the Foundations of Software Engineering (ESEC/FSE ’20)},
  year      = {2020},
  location  = {Virtual Event, USA},
  publisher = {{ACM}},
  doi       = {10.1145/3368089.3417038}
}

@inproceedings{cardenas2009challenges,
  author    = {Alvaro A. C{\'a}rdenas and Saurabh Amin and Bruno Sinopoli and Annarita Giani and Adrian Perrig and Shankar Sastry},
  title     = {Challenges for Securing Cyber‑Physical Systems},
  booktitle = {Workshop on Future Directions in Cyber‑physical Systems Security},
  organization = {DHS},
  address   = {Washington, DC},
  day       = {23},
  month     = {July},
  year      = {2009},
  url       = {http://chess.eecs.berkeley.edu/pubs/601.html}
}

@techreport{anand2024achieving,
  title={Achieving Faster Secure Boot Time on AM26x Devices},
  author={Anand, Nilabh and Kedia, Aakash},
  institution={Texas Instruments},
  year={2024},
  month={November},
  number={SPRADM8},
  type={Application Brief},
  pages={1--5},
  publisher={Texas Instruments Incorporated},
  address={Dallas, Texas},
  url={https://www.ti.com/lit/ab/spradm8/spradm8.pdf},
  note={Application Brief on secure boot optimization techniques for AM263x and AM263Px SoC family}
}

@article{nasser2019accelerated,
  title={Accelerated Secure Boot for Real-Time Embedded Safety Systems},
  author={Nasser, Ahmad M. K. and Gumise, Wonder and Ma, Di},
  journal={SAE International Journal of Transportation Cybersecurity and Privacy},
  volume={2},
  number={1},
  pages={35--48},
  year={2019},
  month={July},
  publisher={SAE International},
  doi={10.4271/11-02-01-0003},
  url={https://doi.org/10.4271/11-02-01-0003},
  issn={1570-761X}
}

@article{sanwald2019secure,
  title={Secure Boot Revisited: Challenges for Secure Implementations in the Automotive Domain},
  author={Sanwald, Steffen and Kaneti, Liron and Stöttinger, Marc and Böhner, Martin},
  journal={SAE International Journal of Transportation Cybersecurity and Privacy},
  volume={2},
  number={2},
  pages={69--81},
  year={2019},
  publisher={SAE International},
  doi={10.4271/11-02-02-0008},
  url={https://doi.org/10.4271/11-02-02-0008},
  note={Paper ID: 11-02-02-0008}
}

@inproceedings{kim2017secure,
  title={Secure Boot Implementation for Hard Real-Time Powertrain System},
  author={Kim, Daehyun and Shin, Eunho and Park, Jin Seo and LEE, KyungSu and Gui, Kok Cheng and Scheibert, Klaus},
  booktitle={SAE World Congress Experience (WCX™ 17)},
  year={2017},
  month={March},
  address={Detroit, Michigan, United States},
  organization={SAE International},
  number={2017-01-1656},
  series={SAE Technical Paper},
  doi={10.4271/2017-01-1656},
  url={https://doi.org/10.4271/2017-01-1656},
  issn={0148-7191},
  note={Paper presented at SAE World Congress Experience, April 4-6, 2017}
}

@phdthesis{kaster2024automotive,
  title={Automotive Software Attestation: Self, Remote, and Peer - Building Trust in Autonomous Driving Safety Systems},
  author={Kaster},
  year={2024},
  school={University of Michigan},
  address={Ann Arbor, Michigan},
  url={https://deepblue.lib.umich.edu/handle/2027.42/192664}
}

@misc{ep3923168b1,
  title={Secure Boot at Shutdown},
  author={Rogala, David Ray},
  howpublished={European Patent},
  number={EP3923168B1},
  year={2023},
  month={March},
  day={08},
  note={European Patent Office. Patent granted March 8, 2023},
  assignee={Harman International Industries, Incorporated},
  address={Stamford, CT, US},
  url={https://patentimages.storage.googleapis.com/f8/10/8d/8afba952324c86/EP3923168B1.pdf},
  filing_date={2021-05-07},
  priority_date={2020-06-10},
  priority_country={US},
  priority_number={202016897918},
  classification={G06F 21/57},
  type={Patent}
}

@article{hoppe2009applying,
  title={Applying intrusion detection to automotive IT-early insights and remaining challenges},
  author={Hoppe, T. and Kiltz, S. and Dittmann, J.},
  journal={Journal of Information Assurance and Security (JIAS)},
  volume={4},
  pages={226--235},
  year={2009},
  month={January},
  publisher={Dynamic Publishers},
  note={Early work on automotive intrusion detection systems}
}

@inproceedings{kwon2018mitigation,
  title={Mitigation mechanism against in-vehicle network intrusion by reconfiguring ECU and disabling attack packet},
  author={Kwon, H. and Lee, S. and Choi, J. and Chung, B. H.},
  booktitle={Proc. of 2018 International Conference on Information Technology (InCIT)},
  pages={1--5},
  year={2018},
  organization={IEEE}
}

@inproceedings{dafoe2024enabling,
  title={Enabling Real-Time Restoration of Compromised ECU Firmware in Connected and Autonomous Vehicles},
  author={Dafoe, J. and Singh, H. and Chen, N. and Chen, B.},
  booktitle={Proc. of Security and Privacy in Cyber-Physical Systems and Smart Vehicles (SmartSP)},
  pages={15--33},
  year={2024},
}

@inproceedings{mansor2015dont,
  title={Don't Brick Your Car: Firmware Confidentiality and Rollback for Vehicles},
  author={Mansor, Hafizah and Markantonakis, Konstantinos and Akram, Raja Naeem and Mayes, Keith},
  booktitle={2015 International Conference on Availability, Reliability and Security (ARES)},
  pages={139--148},
  year={2015},
  month={August},
  organization={IEEE}
}

@inproceedings{gui2018hardware,
  title        = {Hardware Based Root of Trust for Electronic Control Units},
  author       = {Yutian Gui and Ali Shuja Siddiqui and Fareena Saqib},
  booktitle    = {SoutheastCon 2018},
  year         = {2018},
  pages        = {1--7},
  publisher    = {IEEE},
  doi          = {10.1109/SECON.2018.8479266},
  url          = {https://ieeexplore.ieee.org/document/8479266}
}

@inproceedings{kaster2021sliced,
  title={Sliced secure boot},
  subtitle={sampled secure boot with re-usable fingerprints},
  author={Kaster, R. and Ma, D. and Behl, A. and Bakalarczyk, B.},
  booktitle={19th escar Europe: The World's Leading Automotive Cyber Security Conference},
  year={2021},
  month={November}
}

@incollection{dafoe2025hardware,
  title={Hardware-Assisted Runtime In-vehicle ECU Firmware Self-attestation and Self-repair},
  author={Dafoe, Josh and Siy, Job and Chen, Niusen and Chen, Bo},
  booktitle={Proc. of Security and Privacy in Cyber-Physical Systems and Smart Vehicles (SmartSP)},
  pages={140--161},
  year={2025},

}

@article{oham2021bferl,
  title={B-FERL: Blockchain based framework for securing smart vehicles},
  author={Oham, Chuka and Michelin, Regio A. and Kanhere, Salil S. and Jurdak, Raja and Jha, Sanjay K.},
  journal={Information Processing \& Management},
  volume={58},
  number={1},
  pages={102426},
  year={2021},
  publisher={Elsevier},
  doi={10.1016/j.ipm.2020.102426},
  url={https://www.sciencedirect.com/science/article/abs/pii/S0306457320309183},
  note={arXiv preprint arXiv:2007.10528}
}

@inproceedings{khodari2019decentralized,
  title={Decentralized Firmware Attestation for In-Vehicle Networks},
  author={Khodari, Mohammad and Rawat, Abhimanyu and Asplund, Mikael and Gurtov, Andrei},
  booktitle={Proc. of the 5th ACM Cyber-Physical System Security Workshop (CPSS)},
  year={2019},
  publisher={ACM}
}

@misc{nxp_s32g3_2025,
  title={S32G3 vehicle networking reference design},
  author={{NXP Semiconductors}},
  year={2025},
  howpublished={\url{https://www.nxp.com/design/designs/s32g3-vehicle-networking-reference-design:S32G-VNP-RDB3}}
}

@misc{nxp_s32k3_2025,
  title={S32K3 automotive telematics box (T-Box) reference design board},
  author={{NXP Semiconductors}},
  year={2025},
  howpublished={\url{https://www.nxp.com/design/designs/s32k3-automotive-telematics-box-t-box-reference-design-board:S32K3-T-BOX}}
}

@misc{arm_trustzone_cortex_a_2025,
  title={TrustZone for Cortex-A},
  author={{ARM Ltd.}},
  year={2025},
  howpublished={\url{https://www.arm.com/technologies/trustzone-for-cortex-a}},
  note={Accessed: August 3, 2025}
}

@misc{arm_trustzone_cortex_m_2025,
  title={TrustZone for Cortex-M},
  author={{ARM Ltd.}},
  year={2025},
  howpublished={\url{https://www.arm.com/technologies/trustzone-for-cortex-m}},
  note={Accessed: August 3, 2025}
}

@techreport{stmicroelectronics2021an5600,
  title = {{STM32L5} Series {GPIO} Usage with {TrustZone}®},
  author = {{STMicroelectronics}},
  institution = {STMicroelectronics},
  number = {AN5600},
  year = {2021},
  month = {January},
  revision = {1},
  url = {https://www.st.com/resource/en/application_note/an5600-stm32l5-series-gpio-usage-with-trustzone-stmicroelectronics.pdf},
  note = {Application note}
}

@mastersthesis{andreasson2022device,
  title = {Device Attestation for In-Vehicle Network},
  author = {Andr{\'e}asson, Erik and Lyesnukhin, Ivan},
  year = {2022},
  month = {July},
  school = {Chalmers University of Technology and University of Gothenburg},
  address = {Gothenburg, Sweden},
  type = {Master's thesis in Computer science and engineering},
  url = {https://gupea.ub.gu.se/bitstream/handle/2077/74354/CSE%2022-27%20Andr%C3%A9asson%20Lyesnukhin.pdf?sequence=1},
  note = {Master's Thesis 2022, Department of Computer Science and Engineering}
}

@article{kulandaivel2024candid,
  author = {Kulandaivel, S. and Jain, S. and Guajardo, J. and Sekar, V.},
  title = {Candid: A Stealthy Stepping-Stone Attack to Bypass Authentication on {ECU}s},
  journal = {ACM Journal on Autonomous Transportation Systems},
  year = {2024},
  month = {april},
  doi = {10.1145/3657645},
  url = {https://doi.org/10.1145/3657645}
}

@incollection{herrewegen2018beneath,
  author = {den Herrewegen, V. and Garcia, F.},
  editor = {Lopez, P. and Zhou, W. and Soriano, E.},
  title = {Beneath the Bonnet: A Breakdown of Diagnostic Security},
  booktitle = {Computer Security},
  pages = {305--324},
  publisher = {Springer International Publishing},
  address = {Cham},
  year = {2018}
}

@inproceedings{guan2017supporting,
  title={Supporting transparent snapshot for bare-metal malware analysis on mobile devices},
  author={Guan, Le and Jia, Shijie and Chen, Bo and Zhang, Fengwei and Luo, Bo and Lin, Jingqiang and Liu, Peng and Xing, Xinyu and Xia, Luning},
  booktitle={Proceedings of the 33rd annual computer security applications conference},
  pages={339--349},
  year={2017}
}

@inproceedings{lauser2023formal,
  author = {Lauser, T. and Krau{\ss}, C.},
  title = {Formal Security Analysis of Vehicle Diagnostic Protocols},
  booktitle = {Proc. of the 2023 International Conference on Availability, Reliability and Security (ARES)},
  year = {2023},
  publisher = {ACM}
}

@techreport{bielawski2020FirmwareUpdates,
  title        = {Cybersecurity of Firmware Updates},
  author       = {Bielawski, Russ and Gaynier, Ron and Ma, Di and Lauzon, Sam and Weimerskirch, Andre},
  year         = {2020},
  month        = oct,
  institution  = {United States. Department of Transportation. National Highway Traffic Safety Administration},
  number       = {DOT HS 812 807},
  type         = {Final report},
  doi          = {10.21949/1530213},
  url          = {https://rosap.ntl.bts.gov/view/dot/55729}
}

@inproceedings{oguma2008new,
  title={New Attestation-Based Security Architecture for In-Vehicle Communication},
  author={Oguma, Hisashi and Yoshioka, Akira and Nishikawa, Makoto and Shigetomi, Rie and Otsuka, Akira and Imai, Hideki},
  booktitle={IEEE Global Telecommunications Conference (GLOBECOM)},
  pages={1--6},
  year={2008},
  organization={IEEE},
  doi={10.1109/GLOCOM.2008.ECP.369}
}

@inproceedings{kohnhauser2019ensuring,
  title={Ensuring the Safe and Secure Operation of Electronic Control Units in Road Vehicles},
  author={Kohnhäuser, Florian and Püllen, Dominik and Katzenbeisser, Stefan},
  booktitle={2019 IEEE Security and Privacy Workshops (SPW)},
  pages={126--131},
  year={2019},
  organization={IEEE},
  address={San Francisco, CA, USA},
  month={May},
  doi={10.1109/SPW.2019.00033}
}

@inproceedings{hellemans2025spark,
  title     = {SPARK: Secure Privacy-Preserving Anonymous Swarm Attestation for In-Vehicle Networks},
  author    = {Hellemans, Wouter and El Kassem, Nada and Rabbani, Md Masoom and Dushku, Edlira and Chen, Liqun and Braeken, An and Preneel, Bart and Mentens, Nele},
  booktitle = {Proceedings of the 10th IEEE European Symposium on Security and Privacy (EuroS\&P)},
  year      = {2025},
  pages     = {903--917},
  publisher = {IEEE},
  doi       = {10.1109/EuroSP63326.2025.00056}
}

@techreport{autosar_timesync_can_2023,
  title        = {Specification of Time Synchronization over {CAN} (CanTSyn)},
  institution  = {{AUTOSAR} - Automotive Open System Architecture},
  year         = {2023},
  number       = {Document ID 674},
  type         = {Specification},
  address      = {Munich, Germany},
  note         = {{AUTOSAR} Classic Platform Release R23-11},
  url          = {https://www.autosar.org/fileadmin/standards/R23-11/CP/AUTOSAR_CP_SWS_TimeSyncOverCAN.pdf}
}

@manual{nxp2021an13497,
  title        = {AN13497: Firmware Update Using Secondary Bootloader},
  author       = {{NXP Semiconductors}},
  organization = {NXP Semiconductors},
  year         = {2021},
  month        = {December},
  url          = {https://www.nxp.com/docs/en/application-note/AN13497.pdf}
}

@techreport{valasek2015remote,
  author = {Valasek, Chris and Miller, Charlie},
  title = {Remote Exploitation of an Unaltered Passenger Vehicle},
  institution = {IOActive},
  year = {2015}
}

@inproceedings{2023fuzzy,
author = {Pozzobon, Enrico and Weiss, Nils and Mottok, Juergen and Matoušek, Václav},
year = {2023},
month = {10},
pages = {},
title = {Fuzzy fault injection attacks against secure automotive bootloaders},
doi = {10.13154/294-10381}
}

@inproceedings{wan2020rustee,
  title={RusTEE: developing memory-safe ARM TrustZone applications},
  author={Wan, Shengye and Sun, Mingshen and Sun, Kun and Zhang, Ning and He, Xu},
  booktitle={Proceedings of the 36th Annual Computer Security Applications Conference},
  pages={442--453},
  year={2020}
}

@techreport{autosar_fota,
  author      = {{AUTOSAR}},
  title       = {Requirements on Firmware Over-The-Air},
  institution = {AUTOSAR -- Automotive Open System Architecture},
  number      = {Document ID 944},
  year        = {2020},
  month       = nov,
  note        = {AUTOSAR Classic Platform Release R20-11},
  url         = {https://www.autosar.org/fileadmin/standards/R20-11/CP/AUTOSAR_RS_FirmwareOverTheAir.pdf}
}
